%% file: root.tex
\documentclass[letterpaper, 10 pt, conference]{ieeeconf}  % Comment this line out if you need a4paper

\IEEEoverridecommandlockouts                              % This command is only needed if 
\usepackage{graphicx}      % include this line if your document contains figures
\usepackage{tabularx}
\usepackage{amsmath}
\usepackage{booktabs}
\usepackage{multirow}
\usepackage{multicol}   
\usepackage{array}      
\usepackage{siunitx}
\usepackage[table]{xcolor}
\usepackage{hyperref}%%MF Makes references and arXiv links clickable. Very useful! :-)
\usepackage{balance}

\usepackage[backend=biber,style=ieee]{biblatex}
\newcommand{\MSF}{\ensuremath{\mathrm{MSF}}}
\newcommand{\SF}{\ensuremath{\mathrm{SF}}}
\newcommand{\MAD}{\ensuremath{\mathrm{MAD}}}
\newcommand{\TIB}{\mathrm{TIB}}
\newcommand{\TID}{\mathrm{TID}}

\newcommand{\CT}{\mathrm{CT}}
\newcommand{\p}{\mathrm{p}}
\newcommand{\Load}{\mathrm{L}}
\newcommand{\E}{\mathrm{E}}

\title{\LARGE \bf
Multi-Worker Assembly Line Rebalancing with Relevance-Guided Configuration Preservation%*
}

\author{\authorblockN{Martina Vinetti$^*$, Sabino Roselli, Martin Fabian}%
\authorblockA{\textit{Department of Electrical Engineering}\\
\textit{Chalmers University of Technology}\\
Gothenburg, Sweden}%
\thanks{$^*$ Corresponding author, \texttt{\small vinetti@chalmers.se}}%
\thanks{This work was supported by the EUREKA ITEA4 ArtWork project (2023-00970), and the Wallenberg AI, Autonomous Systems and Software Program (WASP) funded by the Knut and Alice Wallenberg Foundation.}%
}

\begin{document}

\maketitle
\thispagestyle{empty}
\pagestyle{empty}
%%%%%%%%%%%%%%%%%%%%%%%%%%%%%%%%%%%%%%%%%%%%%%%%%%%%%%%%%%%%%%%%%%%%%%%%%%%%%%%%
\begin{abstract}

In assembly line balancing, tasks are assigned to stations in order to satisfy a required cycle time. When production conditions change, the line must be rebalanced by modifying the current task allocation, typically aiming to move as few tasks as possible between stations. Similarity measures are commonly used to control such changes, but they generally evaluate configuration preservation by treating all tasks equally, which may not reflect their different practical importance.

In this work, a \emph{pruned Mean Similarity Factor} is proposed for assembly line rebalancing, evaluating similarity only over a subset of structurally relevant tasks identified through a relevance score. The proposed measure is integrated into a compact mixed-integer linear programming (MILP) formulation that considers practical aspects of manual assembly, specifically workload balance, ergonomic exposure, multi-worker stations, and positional constraints.

Computational experiments on extended benchmark instances derived from the literature show that the proposed approach can obtain optimal rebalancing solutions within reasonable computational times, while maintaining high task colocation and balanced workload and ergonomic distributions. In particular, focusing the similarity evaluation on relevant tasks helps reduce the computational effort. 
%%MF required to solve the problem.
\end{abstract}

%%%%%%%%%%%%%%%%%%%%%%%%%%%%%%%%%%%%%%%%%%%%%%%%%%%%%%%%%%%%%%%%%%%%%%%%%%%%%%%%
\section{INTRODUCTION}
\input{Introduction_2}

\section{MODEL DESCRIPTION}
\input{Model_2}

\section{BENCHMARK INSTANCES}
\input{Benchmark}

\section{COMPUTATIONAL EXPERIMENTS AND RESULTS}
\input{Results_2}

\section{CONCLUSIONS}
\input{Conclusions}

%% \addtolength{\textheight}{-12cm}   % This command serves to balance the column lengths
                                  % on the last page of the document manually. It shortens
                                  % the textheight of the last page by a suitable amount.
                                  % This command does not take effect until the next page
                                  % so it should come on the page before the last. Make
                                  % sure that you do not shorten the textheight too much.

%%%%%%%%%%%%%%%%%%%%%%%%%%%%%%%%%%%%%%%%%%%%%%%%%%%%%%%%%%%%%%%%%%%%%%%%%%%%%%%%

%%%%%%%%%%%%%%%%%%%%%%%%%%%%%%%%%%%%%%%%%%%%%%%%%%%%%%%%%%%%%%%%%%%%%%%%%%%%%%%%

%%%%%%%%%%%%%%%%%%%%%%%%%%%%%%%%%%%%%%%%%%%%%%%%%%%%%%%%%%%%%%%%%%%%%%%%%%%%%%%%

\balance
\printbibliography

%%MF \balance

\end{document}

%% file: Introduction_2.tex
Assembly lines are a cornerstone of modern manufacturing, enabling efficient and repeatable production of complex products through a structured sequence of workstations. In such systems, tasks are distributed across stations and executed within a prescribed cycle time that determines the production rhythm. Determining a feasible and efficient allocation of tasks to stations is the subject of the classical Assembly Line Balancing Problem \cite{salveson1955assembly}. Over the past decades, this problem has been extensively studied, leading to numerous variants and solution approaches \cite{becker2006survey, boysen2022survey}.

In real production environments, however, assembly lines rarely remain static over time. Changes in demand, product design, or production targets frequently require adjustments to the existing task allocation. In such situations, redesigning the entire line from scratch is often impractical due to the operational costs associated with operator retraining, relocation of equipment, adjustments to material logistics, and modifications to workstation layouts \cite{gamberini2006, makssoud2015rebalancing, katiraee2023}.
Consequently, large deviations from the current configuration are generally discouraged in practice. 
This setting defines the Assembly Line Rebalancing Problem (ALRBP), in which a new task allocation must be determined while limiting deviations to the existing configuration \cite{cimen2022review}. 

Several approaches have been proposed to control the extent of the changes introduced during rebalancing. Some works explicitly quantify the deviation from the current configuration, for example through similarity measures such as the Mean Similarity Factor ($\MSF$) \cite{gamberini2006}, which evaluates the extent to which tasks assigned to the same station remain colocated after rebalancing. Other approaches limit the number of task reallocations by minimizing \cite{sanci2017} or bounding~\cite{camli2025} the number of tasks that can 
%%MF change station assignments. 
be moved between stations.
A different line of research models rebalancing costs explicitly %%MF, introducing 
by
penalties associated with relocating tasks or modifying station assignments~\cite{gamberini2006}. All these approaches aim to limit or quantify changes while maintaining operational continuity.

However, most existing metrics used to quantify deviations from the current configuration treat all tasks equally, implicitly assuming that preserving the assignment of any task is as important as preserving that of any other. In practice, some tasks have a stronger influence than others on the workstation layout, the feasibility of the line configuration, and the cost of implementing changes. Similar observations have been reported in \cite{mga_rebalancing}, which assign larger relocation penalties to tasks with longer processing times, recognizing their greater impact on the rebalancing process. These considerations suggest that the evaluation of configuration changes should account for the heterogeneous importance of tasks rather than treating them uniformly.

At the same time, many industrial assembly lines, particularly in the 
%%MF production 
final assembly
of large products such as vehicles, rely predominantly on manual operations. In such environments, rebalancing decisions must also consider practical aspects related to workforce organization, work areas, and ergonomics.

One important aspect concerns the presence of stations where multiple workers operate simultaneously on the same product. While this setup enables parallel task execution and can improve productivity, it also introduces additional coordination requirements. Despite its practical relevance, ALRBP with multiple workers per station has received relatively limited attention in the literature. One example is the study by \cite{camli2025}. 

Another relevant aspect concerns positional constraints associated with the physical layout of the product. Workers typically operate at specific mounting locations around the workpiece, and different tasks may require access to distinct areas. If these aspects are not properly considered, multiple workers may require access to the same work area simultaneously or repeatedly cross each other while accessing their assigned work areas. Some studies have begun to incorporate these aspects, for instance by introducing positional constraints to ensure feasible working areas \cite{yang2022} or by analyzing spatial incompatibilities within stations \cite{becker2009}.

Finally, human factors play a central role in manual assembly systems. Therefore, \emph{ergonomic aspects} must be carefully considered, as repetitive motions, awkward postures, and the handling of heavy components may expose operators to musculoskeletal disorders~\cite{ErgRISKS}. For this reason, ergonomic aspects have increasingly been incorporated into assembly line balancing models \cite{otto2011ergonomic, bautista2016} and ALRBP \cite{ErgALRBP}.

Despite the practical relevance of these aspects, the joint consideration of configuration preservation and realistic manual assembly constraints remains relatively limited in the literature. In particular, existing approaches used to evaluate configuration changes rarely account for the heterogeneous importance of tasks in the rebalancing process.

To address %%MF this limitation, 
these limitations,
this work introduces a \emph{pruned Mean Similarity Factor} ($\MSF_\p$), which evaluates configuration similarity by focusing only on a subset of structurally relevant tasks identified through a relevance score. By concentrating the similarity evaluation on tasks that are expected to have a stronger impact on the assembly process, the proposed measure aims to preserve the key structural elements of the current configuration while maintaining flexibility in the reassignment of less influential tasks.

%% This observation motivates the introduction of a similarity measure that differentiates tasks according to their structural relevance. To this end, this work introduces a \emph{pruned Mean Similarity Factor} ($\MSF_\p$), which evaluates configuration similarity by focusing only on a subset of structurally relevant tasks. Task relevance is determined through a relevance score that captures the potential influence of each task on the structure of the assembly process. By concentrating the similarity evaluation on these tasks, the proposed metric aims to preserve the key structural elements of the current configuration while maintaining flexibility in the reassignment of less influential tasks.

In this context, this work investigates the rebalancing of multi-worker assembly lines in manual assembly systems, with the objective of preserving the most structurally relevant elements of the current task allocation while promoting a balanced distribution of workload and ergonomic exposure among workers. To support the practical applicability of the approach while maintaining computational tractability, a compact mixed-integer linear programming (MILP) formulation of the Assembly Line Rebalancing Problem is proposed. Compared with the formulation introduced in \cite{vinetti2025alrbp}, the model adopts a more compact representation aimed at improving scalability.

The proposed approach is evaluated on an extended version of the benchmark dataset~\cite{scholl1993}, to analyze the impact of the proposed similarity measure and assess the scalability of the presented formulation across different problem sizes.

%% Building on these considerations, a new mixed-integer linear programming (MILP) formulation of the ALRBP is proposed. The model integrates a relevance-aware similarity measure, workload balance, ergonomic considerations, and positional constraints in assembly systems with multiple workers per station. Compared with the formulation introduced in~\cite{vinetti2025alrbp}, the proposed model adopts a more compact representation aimed at improving scalability.
%%The proposed approach is evaluated on an extended version of the benchmark dataset~\cite{scholl1993}, to analyze the impact of the proposed similarity measure and assess the scalability of the formulation across different problem sizes.

%% file: Model_2.tex
Consider an assembly line consisting of a set of stations $S$, staffed by workers in the set $W\!$, and responsible for executing a set of indivisible tasks $T$. The line was originally balanced under a previous production plan, but a change in the required cycle time $CT$ makes rebalancing necessary. The goal is to determine a new task allocation that satisfies precedence and positional constraints—modeled here through work areas within each station—while limiting deviations from the current configuration and promoting fair workload and ergonomic distributions among workers. Here, workload 
%%MF denotes 
is measured as the cumulative processing time of the tasks assigned to a worker.

Each task $i \in T$ is characterized by a processing time $\tau_i$, an ergonomic load $e_i$, and a work area requirement $a_i \in A$, indicating the work area in which the task must be performed. Here $A=\{1,2\}$ denotes the set of work areas within a station. For each task $i \in T$, the set $\pi_i$ contains the immediate predecessors of task $i$.

\subsection{Modeling approach}

The proposed formulation builds upon the model introduced in \cite{vinetti2025alrbp}, while adopting a more compact representation aimed at improving scalability. In particular, explicit task--worker and worker--station assignment variables are not included. Since workers are assumed to be identical and interchangeable, modeling worker identities would only increase the number of decision variables without affecting the resulting task allocation.

Workers are instead represented implicitly through the work area structure of each station. Each station can host either one or two workers. When two workers are present, the station is partitioned into two work areas, each associated with one worker, who operates exclusively within that area.
%%Tasks are characterized by a predefined work area attribute indicating the area in which they must be performed.
%%When two workers are present, the station is partitioned into two work areas, each associated with one worker. Tasks are characterized by a predefined work area attribute indicating the area in which they must be performed. Consequently, assigning a task to a station implicitly determines the work area in which it is performed.

This representation allows worker-level quantities to be captured without introducing explicit worker variables. 
%%In stations with two workers, workload and ergonomic exposure are evaluated separately for the two work areas, whereas in single-worker stations they coincide with the total workload and ergonomic load of the station, as the worker can operate across both areas. 
In stations with two workers, worker-level workload and ergonomic exposure are obtained by aggregating the processing times and ergonomic loads of the tasks assigned to each work area. In single-worker stations, the workload and ergonomic exposure of the worker coincide with the total workload and ergonomic load of the station, since the worker can operate across both areas. In this way, worker-level workload and ergonomic measures are modeled while maintaining a compact formulation.

\subsection{Mathematical Formulation}

\noindent\textbf{Sets}

\vspace{1pt}

\noindent
\begin{tabularx}{\linewidth}{lX}
$T$ & set of tasks.\\
$S$ & set of stations.\\
$W$ & set of workers.\\
$A$ & set of work areas within a station.\\
$\pi_i$ & set of immediate predecessors of task $i\in T$.\\
$N_i$ & set of tasks sharing the same station as task $i\in T$ in the current configuration. \\
\end{tabularx}

\vspace{2pt}
\noindent\textbf{Parameters}

\noindent
\begin{tabular}{ll}
$\tau_i$ & processing time of task $i\in T$. \\
$e_i$ & ergonomic load associated with task $i\in T$. \\
$a_i$ & work area required by task $i\in T$, $a_i \in A$. \\
$\CT$ & cycle time. \\
\end{tabular}

\vspace{3pt}
\noindent\textbf{Decision Variables}

\vspace{0pt}
%\medskip
\noindent\emph{Binary variables}

\noindent
\begin{tabularx}{\linewidth}{lX}
$x_{is}$   & 1 if task $i$ is assigned to station $s$, 0 otherwise.\\
$u_s$ & 1 if station $s$ has one worker, 0 if it has two.\\
\end{tabularx}

\vspace{1pt}
\noindent\emph{Auxiliary continuous variables}

\noindent
\begin{tabularx}{\linewidth}{lX}
$L_{sa}$ & workload of work area $a$ at station $s$.\\
$L_s$ & total workload at station $s$.\\
$E_{sa}$ & ergonomic load of work area $a$ at station $s$.\\
$E_s$ & total ergonomic load at station $s$.\\
\end{tabularx}

\subsection{Fairness measures}

Let $L_w$ and $E_w$ denote the workload and ergonomic load associated with worker $w \in W$. These quantities are derived from the station and work area loads defined in the model. In two-worker stations, each worker corresponds to one work area, while in single-worker stations both quantities coincide with the total workload and ergonomic load of the station.

The average workload and ergonomic load are defined as the total workload and ergonomic load divided by the number of workers:
\begin{align}
\overline{\!L} = \frac{\sum_{i\in T}\tau_i}{|W|},
\\
\overline{\!E} = \frac{\sum_{i\in T}e_i}{|W|}.
\end{align}

Workload and ergonomic fairness are evaluated through the mean absolute deviation:
\begin{align}
\MAD_\Load &= \frac{1}{|W|}\sum_{w\in W}\left|L_w-\,\overline{\!L}\,\right|, \\
\MAD_\E &= \frac{1}{|W|}\sum_{w\in W}\left|E_w-\,\overline{\!E}\,\right|.
\end{align}

In the MILP formulation, the absolute-value expressions are handled through standard linearization.

\subsection{Similarity measure}

To preserve the current task allocation during rebalancing, a widely used measure is the Mean Similarity Factor (\MSF). For each task $i$, the corresponding similarity factor $\SF_i$ measures how many of the tasks previously assigned to the same station remain colocated after rebalancing. The classical \MSF\ \cite{gamberini2006} is defined as:
\begin{equation}
\MSF = \frac{1}{|T|}\sum_{i\in T}\SF_i,
\end{equation}
with
\begin{equation}
\SF_i = \frac{|\TIB_i \cap \TID_i|}{|\TIB_i|},
\end{equation}
where $\TIB_i$ and $\TID_i$ denote the sets of tasks, excluding $i$, assigned to the same station as task $i$ in the current and rebalanced configurations, respectively.

By construction, the classical \MSF\ treats all tasks uniformly.
%%MF In practice, some tasks have a stronger impact on station layout, line feasibility, or the cost of implementing changes, for example because they require specific tools, resources, or working conditions, or because they occupy structurally critical positions in the assembly process. 
However, in practice some tasks have a stronger impact on station layout, line feasibility, or the cost of implementing changes.
This could be due to requiring specific tools, resources, or working conditions, or because they occupy structurally critical positions in the assembly process.
Preserving the colocations of these tasks is often more important than preserving those of less influential ones.

For this reason, a pruned similarity metric is introduced. Let $R\subseteq T$ denote the set of \emph{relevant} tasks. The pruned Mean Similarity Factor is defined as:
\begin{equation}
\MSF_\p = \frac{1}{|R|}\sum_{i\in R}\SF_i.
\end{equation}

%%%%%%%%%%%%%%%%%%%%
%%The set $R$ is obtained by ranking tasks according to a relevance score and selecting the top $K\,\%$ of tasks, where $K$ is a pruning parameter controlling the proportion of tasks included in the similarity evaluation. For each task $i\in T$, the relevance score is:
%\begin{equation}
%\rho_i = \tilde{\tau}_i + \tilde{e}_i + \tilde{d}_i,
%\end{equation}
%where $\tilde{\tau}_i$, $\tilde{e}_i$, and $\tilde{d}_i$ denote the normalized processing time, ergonomic load, and structural importance of task $i$, respectively. The latter is computed from the degree of node $i$ in the precedence graph, i.e., from the number of immediate predecessors and successors. 
%These three terms capture complementary aspects that influence the impact of a task on the rebalancing process. Tasks with longer processing times typically have a stronger effect on workload distribution across stations, while tasks with high ergonomic load contribute significantly to the ergonomic balance among workers. The structural term $\tilde{d}_i$ reflects the position of the task in the precedence graph; tasks with many predecessors or successors constrain the placement of several other tasks and therefore play a key role in the assembly sequence. 

%In this way, the similarity evaluation focuses on tasks that are expected to have the largest impact on the structure of the rebalanced solution.
%%%%%%%%%%%%%%%%%%%%%%
The set $R$ is obtained by ranking tasks according to a relevance score and selecting the top $K\,\%$ of tasks, where $K$ is a pruning parameter controlling the proportion of tasks included in the similarity evaluation. 
The choice of $K$ reflects the desired level of configuration preservation, with larger values prioritizing the retention of the current task allocation and smaller values allowing greater flexibility during rebalancing.
The relevance score associated with each task $i \in T$ is defined as:
\begin{equation}
\rho_i = \tilde{\tau}_i + \tilde{e}_i + \tilde{d}_i,
\end{equation}
where $\tilde{\tau}_i$, $\tilde{e}_i$, and $\tilde{d}_i$ denote the normalized processing time, ergonomic load, and structural importance of task $i$, respectively. The latter is computed from the degree of node $i$ in the precedence graph, i.e., from the number of immediate predecessors and successors.
The proposed relevance score combines processing time, ergonomic load, and structural importance in the precedence graph. These factors are used here as indicators of task relevance, although alternative or additional task attributes may be incorporated depending on the characteristics of the assembly system.

The similarity evaluation is then restricted to the highest-ranked tasks, focusing configuration preservation on the most relevant elements of the current allocation.

\subsection{Objective function}

To combine the different components in a single objective, the fairness terms are normalized to obtain quantities on comparable scales. This avoids the objective being dominated by a component solely due to its numerical magnitude. After normalization, the three objective components are aggregated with equal weights, reflecting the absence of a priori preferences among workload fairness, ergonomic fairness, and configuration preservation. Let
%%To combine the different components in a single objective, the fairness terms are normalized. Let
\begin{align}
\overline{\MAD}_\Load = \frac{\MAD_\Load}{\CT},
\\
\overline{\MAD}_\E = \frac{\MAD_\E}{\sum_{i\in T} e_i}.
\end{align}
%%MF
The objective function is then defined as
\begin{equation}
\min \; \overline{\MAD}_\Load + \overline{\MAD}_\E + (1 - \MSF_\p).
\end{equation}

\subsection{Constraints}

Each task is assigned to exactly one station,
\begin{equation}
\sum_{s\in S} x_{is} = 1 \qquad \forall i\in T,
\end{equation}
and precedence is enforced through station ordering,
\begin{equation}
\sum_{s\in S} s\,x_{js} \le \sum_{s\in S} s\,x_{is}
\qquad \forall i\in T,\; \forall j\in \pi_i.
\end{equation}

Station workloads are computed separately for each work area and then aggregated:
\begin{subequations}
\begin{align}
L_{sa} &= \sum_{i\in T:\,a_i=a}\tau_i x_{is}
& \forall s\in S,\; a\in A,\\
L_s &= \sum_{a\in A} L_{sa}
& \forall s\in S.
\end{align}
\end{subequations}
The ergonomic loads are defined analogously:
\begin{subequations}
\begin{align}
E_{sa} &= \sum_{i\in T:\,a_i=a} e_i x_{is}
&\forall s\in S,\; a\in A,\\
E_s &= \sum_{a\in A} E_{sa}
&\forall s\in S.
\end{align}
\end{subequations}

These quantities correspond to the loads associated with the work areas of each station, which represent the workers when a station operates with two workers.

Cycle-time feasibility is enforced at both station and work area level:
\begin{subequations}
\begin{align}
L_s &\le \CT(2-u_s),
& \forall s\in S,\; \label{eq:capacity_total}\\
L_{sa} &\le \CT(1+u_s)
& \forall s\in S,\; a\in A.
\end{align}
\end{subequations}

When $u_s=0$, station $s$ hosts two workers and each work area is individually limited by the cycle time. When $u_s=1$, station $s$ hosts a single worker and only the total station capacity~\eqref{eq:capacity_total} is binding.

%% file: Benchmark.tex
The computational study is based on benchmark instances derived from a widely used public dataset for the assembly line balancing problem~\cite{scholl1993}. Among the 25 available instances, 21 are considered, namely all those with fewer than 100 tasks, resulting in a benchmark with problem sizes ranging from 7 to 94 tasks. Since the original dataset provides task processing times and precedence relations only, two additional task attributes are generated: ergonomic load and work area. 
%%The ergonomic load associated with each task is generated as a synthetic quantity loosely related to task duration, while introducing controlled stochastic variability and occasional high-load tasks. As a result, the generated ergonomic loads are not deterministically correlated with processing times, allowing the two quantities to capture different aspects of task difficulty. 

%%%%%%%%%%%%%%%%%%
%The ergonomic load associated with each task is generated as a synthetic quantity loosely related to task duration while introducing controlled stochastic variability and occasional high-load tasks. For each task $i$, the ergonomic load is defined as
%\begin{equation}
%e_i = \alpha \tau_i + \sigma_\tau U_i \left( 1 + S_i \left(M_i - 1\right) \right),  
%\end{equation}
%%
%where $\tau_i$ denotes the processing time of task $i$ and $\sigma_\tau$ is the standard deviation of the processing times of all tasks in the instance. The random variable $U_i$ introduces bounded stochastic variability around the baseline ergonomic load, while the binary variable $S_i$ models high-load conditions, whose magnitude is controlled by the factor $M_i$. 

%This formulation maintains a positive association between processing time and ergonomic load while allowing heterogeneous ergonomic demands across tasks.
%%%%%%%%%%%%%%%%%%
The ergonomic load associated with each task is generated as a synthetic quantity loosely related to processing time while introducing controlled stochastic variability across tasks. In industrial applications, ergonomic load values could be obtained from established assessment methods or from digital human modeling tools. Here, synthetic scores are generated solely for benchmarking purposes. For each task $i$, the ergonomic load is defined as

\begin{equation}
e_i = \alpha \tau_i + \sigma_\tau U_i \left( 1 + S_i \left(M_i - 1\right) \right),
\end{equation}

where $\tau_i$ denotes the processing time of task $i$ and $\sigma_\tau$ is the standard deviation of the processing times of all tasks in the instance. The random variable $U_i$ introduces bounded stochastic variability around the baseline ergonomic load, while the binary variable $S_i$ models high-load conditions, whose magnitude is controlled by the factor $M_i$.

This formulation maintains a positive association between processing time and ergonomic load while allowing heterogeneous ergonomic demands across tasks.
%%%%%%%%%%%%%%%%%%%%%%%%%%
Each task is assigned a work area attribute $a_i \in A$, indicating the work area in which it must be performed within a station. The assignment is generated by exploiting the topological structure of the precedence graph to obtain spatially coherent task groups, while introducing a limited random perturbation to avoid an artificially sharp separation between areas. %%MF\footnote{The repository containing code and experimental instances is omitted to preserve author anonymity and will be released if the paper is accepted.}

%%MF I put this footnote at the beginning of the experimental section
%%\footnote{Code and experimental instances are available at: \url{https://github.com/Chalmers-Control-Automation-Mechatronics/ALRP}}.

Since the rebalancing problem evaluates the similarity with respect to the current task allocation, a current task allocation must be specified. To obtain such a configuration, a balancing problem is solved for each instance under an initial cycle time, yielding a feasible assignment of tasks to stations. The number of stations is given, and the balancing model corresponds to the proposed ALRBP formulation without $\MSF_\p$ in the objective function. During this preliminary phase, the number of workers is set to the minimum value that ensures feasibility. Since the purpose of this step is simply to obtain a feasible baseline configuration, the balancing problems are not necessarily solved to optimality. The resulting task allocations are then used as the starting configurations for the rebalancing phase, where the line is optimized under a new cycle time, $\CT$, while preserving the same number of stations and again selecting the minimum feasible number of workers.

%% file: Results_2.tex
The computational study investigates two main aspects of the proposed approach. First, the model is evaluated on benchmark instances of increasing size to assess its overall computational performance. Second, the influence of the pruning parameter $K$ is analyzed to assess how the size of the relevant task set $R$ affects both solution quality and computational performance.

All experiments\footnote{Code and experimental instances are available at: \url{https://github.com/Chalmers-Control-Automation-Mechatronics/ALRP}.} were conducted using the commercial MILP solver \emph{Gurobi} (version~12.0.2)~\cite{gurobi}. 
Computations were performed on an Apple Mac Studio equipped with an Apple M4 Max processor and 64\,GB RAM running macOS~Tahoe~26.1. 
A time limit of 5400 seconds was imposed for each instance.

Table~\ref{tab:runtime_summary} summarizes the computational results obtained for the benchmark instances grouped into three size classes according to the number of tasks. For each class, the table reports the number of instances, the average runtime (in seconds), the average values of the objective components, and the number of instances solved to optimality within the time limit. Runtime values are averaged only over instances solved to optimality. The fairness metrics are reported as percentage mean absolute deviations with respect to the average worker workload $\overline{L}$ and average worker ergonomic load $\overline{E}$, denoted by $\MAD^{\%}_{\Load}$ and $\MAD^{\%}_{\E}$, respectively. In the experiments, the pruning parameter is fixed to $K=30\%$, which is used as a representative pruning level. This value was selected based on the results of the preliminary analysis, which indicate that moderate pruning levels maintain high similarity while keeping the computational effort relatively low.

\begin{table}[t]
\centering
\renewcommand{\arraystretch}{1.15}
\setlength{\tabcolsep}{3.5pt}
\caption{Summary of computational results across instance size classes.}
\begin{tabular}{l c c c c c c}
\toprule
\textbf{Size class} & \textbf{\#Inst.} & \textbf{Runtime} & \textbf{$\mathbf{MSF}_\p$} & \textbf{$\mathbf{MAD}^{\%}_\E$} & \textbf{$\mathbf{MAD}^{\%}_\Load$} & \textbf{\#Opt.} \\
\midrule
7--30 tasks   & 10 & 0.36   & 0.80 & 19.19 & 12.24 & 10 \\
31--60 tasks  & 5  & 38.13  & 0.90 & 15.53 & 2.87  & 5  \\
61--94 tasks & 6  & 15.82  & 0.91 & 15.24 & 6.36  & 4  \\
\bottomrule
\end{tabular}
\label{tab:runtime_summary}\vspace{-1em}
\end{table}

The results show that the proposed formulation scales well with the size of the instances. Instances with up to 30 tasks are solved almost instantaneously, while for the class with 31--60 tasks the average runtime remains within tens of seconds. For the largest class (61--94 tasks), the average runtime over the instances solved to optimality is approximately 16 seconds, although two instances in this class reach the time limit. Regarding solution quality, the $\MSF_\p$ remains consistently high across all instance classes, indicating that the rebalanced solutions preserve most of the current task allocation even as the problem size grows.

\begin{figure}[h]
    \centering
    \includegraphics[width=0.95\linewidth]{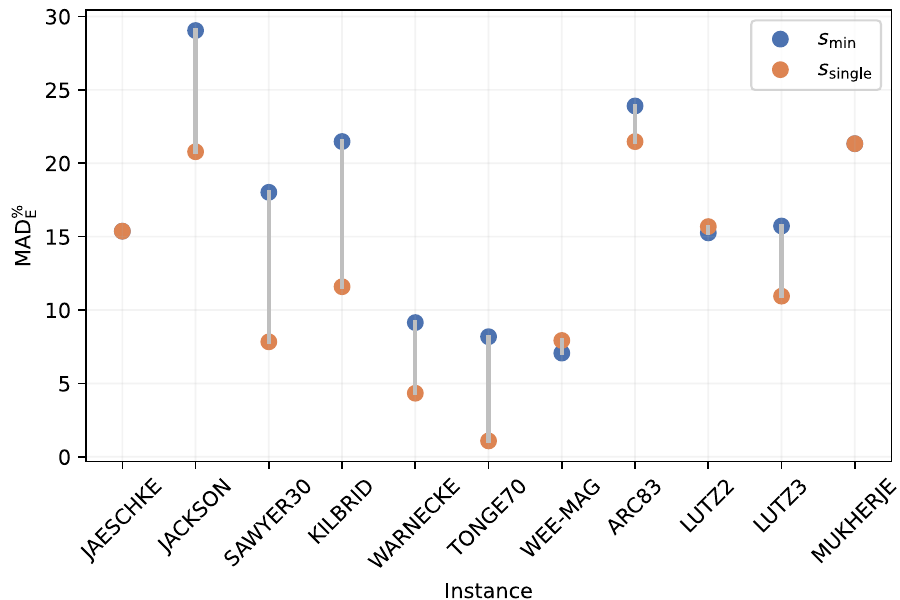}
    \caption{Comparison of $\MAD^{\%}_{\E}$ under two station configurations. Each segment connects the values obtained with the minimum number of stations ($s_{\text{min}}$) and with the configuration allowing one worker per station ($s_{\text{single}}$).}
    \label{fig:erg_stations}
\end{figure}

Both $\MAD^{\%}_{\Load}$ and $\MAD^{\%}_{\E}$ slightly decrease as the instance size increases, since larger instances provide a richer combinatorial assignment space and greater flexibility in redistributing tasks across stations. However, the values of $\MAD^{\%}_{\E}$ remain noticeably higher than those of $\MAD^{\%}_\Load$. One possible explanation for this behavior lies in the structure of the model; workload is implicitly constrained by the station cycle-time limits, whereas ergonomic exposure is not subject to comparable feasibility constraints, making it intrinsically more difficult to balance. To further explore factors that may influence $\MAD^{\%}_{\E}$, Fig.~\ref{fig:erg_stations} reports the $\MAD^{\%}_{\E}$ obtained under two alternative station configurations for a subset of 11 instances.
For each instance, the figure compares the value of $\MAD^{\%}_{\E}$ obtained with the minimum feasible number of stations ($s_{\text{min}}$) and with an extended configuration ($s_{\text{single}}$) that allows at most one worker per station. The station configuration was also analyzed with respect to $\MAD^{\%}_\Load$ and $\MSF_\p$, but no consistent trends were observed. Overall, Fig.~\ref{fig:erg_stations} suggests a tendency towards improved $\MAD^{\%}_{\E}$ when more stations are available. In the $s_{\text{single}}$ configuration, where each station hosts a single worker, the work-area constraints arising in multi-worker stations are relaxed, increasing flexibility within each station and facilitating the balancing of ergonomic exposure.
%%In the $s_{\text{single}}$ configuration, the work-area constraints arising in multi-worker stations are relaxed. This increases the flexibility of task allocation within each station, facilitating the balancing of ergonomic exposure without necessarily inducing additional task reallocations across stations.
%%For each instance, the figure compares the value of $\MAD_\E$ obtained with the minimum feasible number of stations ($s_{\min}$) and with an extended configuration ($s_{\text{single}}$) that allows assigning at most one worker per station. The vertical segments therefore illustrate how $\MAD_\E$ changes when the number of stations increases. The station configuration was also analyzed with respect to $\MAD_\Load$ and $\MSF_\p$, but no consistent trends were observed. The results indicate a tendency towards improved $\MAD_\E$ when more stations are available. This behavior can be explained by the structure of the model. When more stations are available, it becomes possible to operate with single-worker stations, which relaxes the work-area constraints that arise in multi-worker stations. As a result, the optimizer gains additional flexibility in redistributing tasks, facilitating a more balanced allocation of ergonomic exposure.

\begin{figure}[h]
    \centering
    \includegraphics[width=0.95\linewidth]{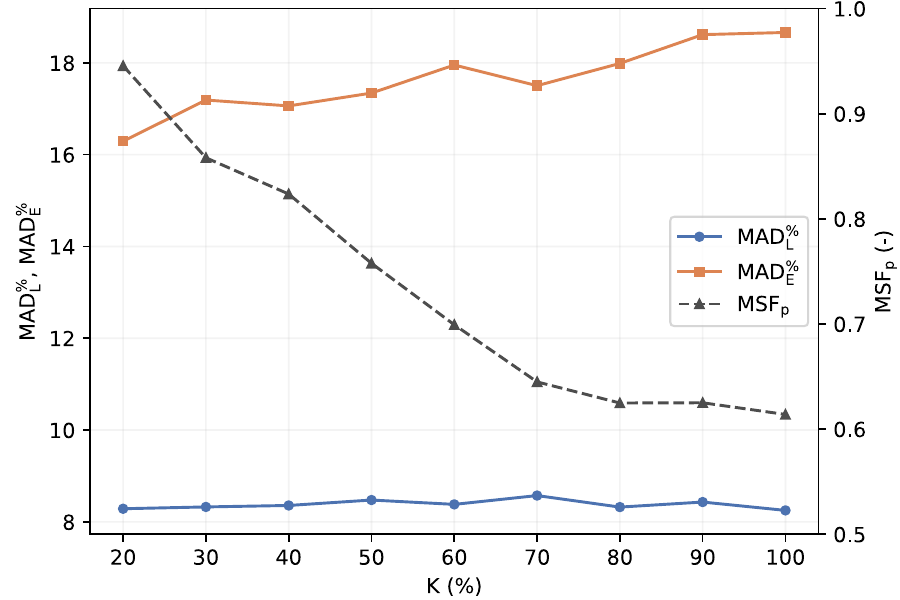}
    \caption{Impact of the pruning level $K$ on $\MAD^{\%}_\Load$, $\MAD^{\%}_\E$ and $\MSF_\p$.}\vspace{-1em}
    \label{fig:solK}
\end{figure}

The following experiments focus on the impact of the similarity measure $\MSF_\p$ by considering different $K$. Fig.~\ref{fig:solK} reports the average values of the three objective components across the 21 benchmark instances for different values of the pruning parameter $K$. 
Values of $K$ smaller than $20\%$ were not considered since $K\leq 10\,\%$ relies on too few tasks and thus approaches the balancing problem rather than being a genuine rebalancing problem. 

The results show that workload fairness is unaffected by the choice of $K$. The average value of $\MAD^{\%}_\Load$ remains consistently low for all tested values of $K$. Ergonomic fairness exhibits a slightly stronger dependence on $K$. In particular, smaller values of $K$ lead to marginally lower values of $\MAD^{\%}_\E$, indicating a modest improvement in the distribution of ergonomic exposure. The behavior of $\MSF_\p$ is consistent with expectations, lower values of $K$ lead to higher similarity between the current and rebalanced task allocation. In particular, the average value of $\MSF_\p$ approaches $0.95$ for $K=20\%$, while it decreases to approximately $0.6$ when $K=100\%$, i.e., when the proposed metric reduces to the classical $\MSF$. When $K$ increases, the similarity measure progressively includes a larger number of tasks, making it more difficult for the model to preserve all original station groupings while simultaneously satisfying the constraints. As a result, deviations from the original configuration become more likely. 
Overall, the results in Fig.~\ref{fig:solK} indicate that evaluating similarity on smaller subsets of structurally relevant tasks effectively preserves the key structural elements of the current task allocation, while allowing less influential tasks to be relocated when necessary to ensure workload or ergonomic balance.

\begin{figure}[h]
    \centering
    \includegraphics[width=0.95\linewidth]{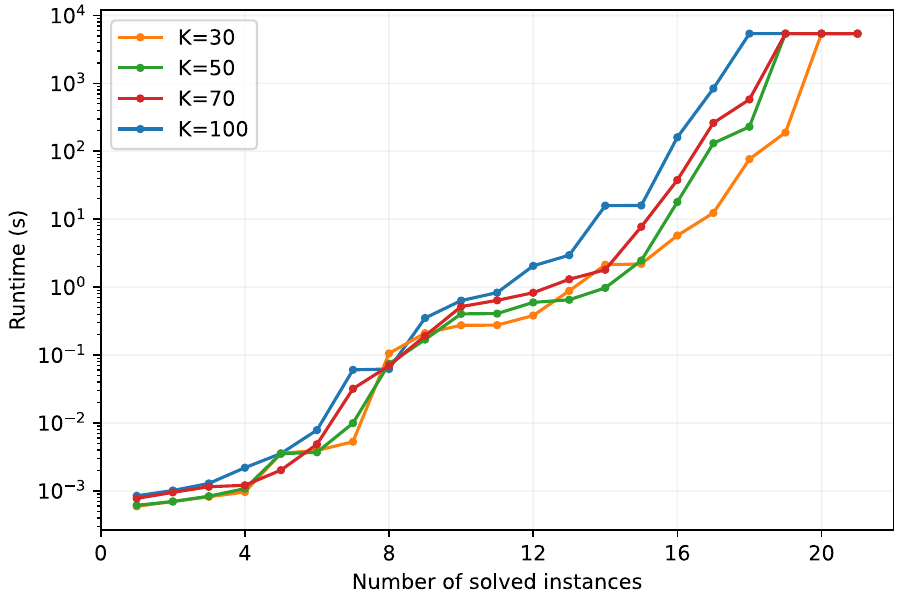}
    \caption{Cactus plot of runtime for different pruning levels $K$.}
    \label{fig:cactusPlot}
\end{figure}

Fig.~\ref{fig:cactusPlot} shows the runtime for different values of $K$ in a cactus plot, enabling a cumulative comparison of computational performance across the dataset.
%%%
%%%
A logarithmic scale is adopted on the vertical axis since the runtime spans several orders of magnitude across the benchmark instances. 
%%This reflects the heterogeneous difficulty of the problems, where many instances are solved within fractions of a second, while the most challenging ones require substantially longer computational times.
%%
The influence of $K$ on the runtime is visible across the entire plot, but becomes more pronounced in the rightmost part, corresponding to the most difficult instances. Smaller values of $K$ generally lead to shorter runtimes, while the differences are less significant for the easier instances. This behavior is consistent with the role of the pruning strategy; when the similarity measure is evaluated on a smaller subset of tasks, the model has greater flexibility in exploring feasible assignments, which reduces the computational effort required to prove optimality for the most challenging problems.

%% file: Conclusions.tex
This work addresses a multi-worker Assembly Line Rebalancing Problem in manual assembly systems, to determine a new task allocation that maintains the colocation of relevant tasks while promoting a fair distribution of workload and ergonomic exposure among workers.

A pruned Mean Similarity Factor ($\MSF_\p$) is introduced to evaluate configuration similarity by focusing on a subset of structurally relevant tasks. The definition of relevance can be adapted to the characteristics of the assembly system.

A compact mixed-integer linear programming formulation of the problem is proposed, incorporating practical aspects of manual assembly systems such as positional constraints, to avoid impractical task assignments in multi-worker stations. Computational experiments on extended benchmark instances show that the proposed formulation can find optimal task allocation within reasonable computational times. The $\MSF_\p$ proves effective in guiding the rebalancing process, enabling solutions that better retain relevant task groupings from the current configuration while achieving a balanced distribution of workload and ergonomic exposure.

Future research will investigate extensions with additional work areas and larger numbers of workers per station, as well as approaches to further improve scalability and adapt the relevance and pruning mechanisms to different industrial requirements.